\documentclass[sigconf]{acmart}
\usepackage{balance}

\AtBeginDocument{%
  \providecommand\BibTeX{{%
    \normalfont B\kern-0.5em{\scshape i\kern-0.25em b}\kern-0.8em\TeX}}}

\setcopyright{acmcopyright}
\copyrightyear{2023} 
\acmYear{2023} 
\setcopyright{acmlicensed}\acmConference[SUMAC '23]{Proceedings of the 5th Workshop on the analySis, Understanding and proMotion of heritAge Contents}{November 2, 2023}{Ottawa, ON, Canada}
\acmBooktitle{Proceedings of the 5th Workshop on the analySis, Understanding and proMotion of heritAge Contents (SUMAC '23), November 2, 2023, Ottawa, ON, Canada}
\acmPrice{15.00}
\acmDOI{10.1145/3607542.3617355}
\acmISBN{979-8-4007-0279-2/23/11}

\begin{document}

\title{Latent Wander: an Alternative Interface for Interactive and Serendipitous Discovery of Large AV Archives}


\author{Yuchen Yang}
\affiliation{%
  \institution{EPFL}
  \city{Lausanne}
  \country{Switzerland}}
\email{yuchen.yang@epfl.ch}

\author{Linyida Zhang}
\affiliation{%
  \institution{EPFL}
  \city{Lausanne}
  \country{Switzerland}}
\email{linyida.zhang@epfl.ch}

\renewcommand{\shortauthors}{Yuchen Yang \& Linyida Zhang}

\begin{abstract}
Audiovisual (AV) archives are invaluable for holistically preserving the past. Unlike other forms, AV archives can be difficult to explore. This is not only because of its complex modality and sheer volume but also the lack of appropriate interfaces beyond keyword search. The recent rise in text-to-video retrieval tasks in computer science opens the gate to accessing AV content more naturally and semantically, able to map natural language descriptive sentences to matching videos. However, applications of this model are rarely seen. The contribution of this work is threefold. First, working with RTS (Télévision Suisse Romande), we identified the key blockers in a real archive for implementing such models. We built a functioning pipeline for encoding raw archive videos to the text-to-video feature vectors. Second, we designed and verified a method to encode and retrieve videos using emotionally abundant descriptions not supported in the original model. Third, we proposed an initial prototype for immersive and interactive exploration of AV archives in a latent space based on the previously mentioned encoding of videos.
\end{abstract}




\maketitle

\section{Introduction}

\subsection{Background}
With multimodality and temporality, audiovisual (AV) content can be documented and preserved in a way that other traditional media cannot. It can document beyond the static and visual perspective, such as emotional changes, relationships between people, and touches of sarcasm. Compared to others, audiovisual content is closer to the complexity of reality, captures individual and collective memories, feelings and histories, and mirrors cultures and aesthetics holistically.

People have gone beyond the traditional keyword-based search to better assist access to such content. Multiple ways of encoding the video content into different types of data and ways to explore and retrieve are invited. At the forefront of these attempts is the recent text-to-video retrieval task. 

A distinguishment between the text-to-video retrieval in this paper and the traditional keyword search is necessary. While the traditional still relies on text or keyword matching for many manual or automated extracted features, the text-to-video retrieval tasks enable a transformer-based text description and video co-embedding and switch the matching of keywords to vector-based similarity search on the holistic semantics. 

\subsection{Motivation}
\begin{figure}
    \centering 
    \includegraphics[width=1\columnwidth]{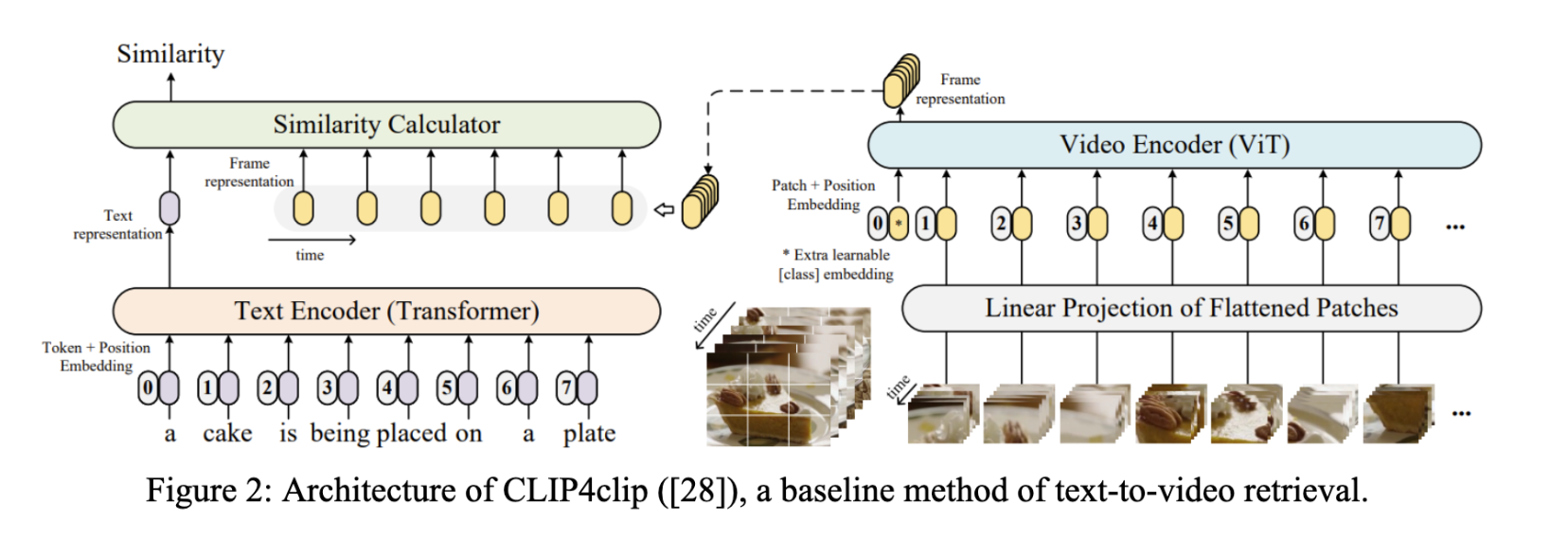} 
    \caption{Architecture of CLIP4clip (\cite{luo2022clip4clip}), a baseline method of text-to-video retrieval.}
\label{2}
\end{figure}

Text-to-video retrieval highlights an encoding and similarity learning task. At a high level, video clips are encoded as vectors in high-dimensional spaces using different encoders for different modalities. When a user inputs a query, the model encodes it into a vector and computes its similarity scores against all video vectors. The retrieval results are the most similar. In the training process, the objective is a latent space where the distance between the paired text and video vectors is minimised. Figure \ref{2} shows the typical structure of text-to-video retrieval models. 

Promising as it sounds, when implemented into real-world applications as the one we worked with, there are several problems yet to be solved: 

(1) Almost all existing pre-trained models perform poorly in the wild. Taking Radio Télévision Suisse (RTS) as an example, there are no useful video content descriptions for such training tasks. Even if a video archive has been annotated, it is of poor quality and not used for training such models \cite{li2022taking, wang2021emotion}...

(2) Manual labels are useful, but humans tend to focus naturally on the most obvious visual perspectives of the video. Even with multiple annotators, descriptions can be partially focused and repetitive \cite{chen2021msr}. More subtle but equally important information, like gestures, colours, lines and more abstract emotions, is often lost in the annotation process. 

(3) There are few applications for such models other than a traditional search box. As archives actively seek more public-engaging ways to allow serendipitous explorations of the archives, a better way to access the archive is yet to be explored.

\subsection{Contribution}
Focusing on the real-world AV archive from RTS, this work will:

1) Design and verify a working pipeline to pre-process real-world archival videos to be able to adapt the text-to-video embedding. 

2) Verify that using the text-to-video retrieval modes, it is possible to go beyond visually focused queries to handle more complex queries with emotions. 

3) As an exemplary application for utilising such a higher-level encoding of videos for serendipitous exploration of large archives.

\section{Related Work}
\subsection{Text-to-Video Retrieval}
The common design of text-to-video retrieval is as follows: a text encoder, which converts the query input into a vector; a video encoder, which extracts multiple features from videos; and a multimodal fusion module, which maps both text features and visual features into a high-dimension space to serve downstream tasks. 

One of the main focuses for progressing the text-to-video models is on obtaining better video representations. VideoBERT \cite{sun2019videobert} is one typical work that includes large-scale pretraining to obtain more detailed information from the video side. Some other models work on finding better ways to deal with the text-video connection. ClipBERT \cite{lei2021less} uses mean pooling on the top of ResNet \cite{he2016deep} to gather temporal information, and Frozen \cite{bain2021frozen} adopts a contrastive learning way to learn video-text relations.

In recent years, video encoding methods have started to reply to multimodalities of the video and incorporate more information in the process. \cite{gabeur2020multi} is one of the earliest to attempt to obtain more information from the video outside of the visual modality, introducing other cues like the audio to the game.

With the active progress made on the video encoding side, one can safely say that video encoding contains very abundant and multimodal information. However, there are very few works taking on the text descriptions side. With most text descriptions being plain, short and visually focused, the information on the two sides of the joint embedding becomes unbalanced. \cite{chen2021msr} very briefly examined how text annotation quality is tied to text-to-video models' performance. 

\subsection{Explorative Interfaces for Archival Content}
Accessibility to the public, especially in the context of cultural and heritage institutions,  has been reframed from a simple search or linear narrative to responsive, immersive, and personalised \cite{janes2019museum}. This drives the wave of probes into digital technologies and interactive, explorative, and embodied experiences for providing access to the public. Instead of focusing on curated exhibitions and a simple search box, institutions introduce experimental approaches that "open up" the full potential of the archival content and allow viewers to interact, explore, and start to create their personalised understanding of the content \cite{kenderdine2020prosthetic, brown2000interactive}.

The National Museum of Film in the Netherlands created the "Film Catcher" \cite{eyecatcher}, which allows viewers to search 'visual' but not film title or director. By automatically extracting these not-often-used features, this playful, intuitive, explorative interface allows users to build their own experiences of browsing and watching and discover serendipitous connections between historical and cultural AV content. Similar to this, the University of Amsterdam research project "The Sensory Moving Image Archive" \cite{masson2020exploring} has explored, in a more theoretical way, how the use of sensory features such as colour and movement, combined with possibilities for explorative browsing, would provide a boost to the practice of those users who seek to repurpose collections creatively.

In this direction, text, image, music, and even AV content can be encoded into some holistic higher-level feature vectors \cite{choi2020encoding, gabeur2020multi, zhang2021multi, yang2022tableformer}. These vectors are essential for building connections of content in a novel way. Surprise Machines \cite{rodighiero2022surprise} from Harvard Metalab is a visual investigation that sets out to analyse, connect and visualize the entire universe of museums’ collections using such a latent space. The objective is to open up unexpected insights into the more than 200,000 images previously strictly catalogued by archivists. 

\section{The RTS} 
This project aims to test and verify designed pipelines and improved methods in real-world settings. For the ease of this work, a subset of the whole RTS archive is chosen. We select a random sample of videos with “TypeContenu - TV series and films" for comprehensiveness, as they contain more complex narrative content than news. The final dataset used for this work contains around 1,000 videos with a total runtime of around 530 hours. This yields an average runtime for each video of 0.53 hours, with no textual descriptions of the video. Standard datasets used for Text-to-video retrieval tasks, such as MSR-VTT, have an average of 15 seconds in video runtime. For each video, 20 corresponding textual descriptions are attached. The real-world archive would require intensive pre-processing to adapt to state-of-the-art encoding techniques.

\section{Methods}
The overall pipeline is designed as Figure \ref{3}. The pipeline's upper part (outside the dashed box) focuses on pre-processing the RTS or any other real-world dataset into a useful format and structure. The later part of the pipeline consists of two separate strategies to handle the query with added emotions. As depicted in Figure \ref{4}, one strategy requires changes in the training set preparation, where all descriptions are rewritten to include emotional factors; the other strategy is to use the original text-to-retrieval model and training data but split the query into two steps, one handling the natural language description query, one use binary search on emotional tags in the form of metadata. Each important component of this pipeline is elaborated as follows:
\begin{figure}
    \centering 
    \includegraphics[width=.4\columnwidth]{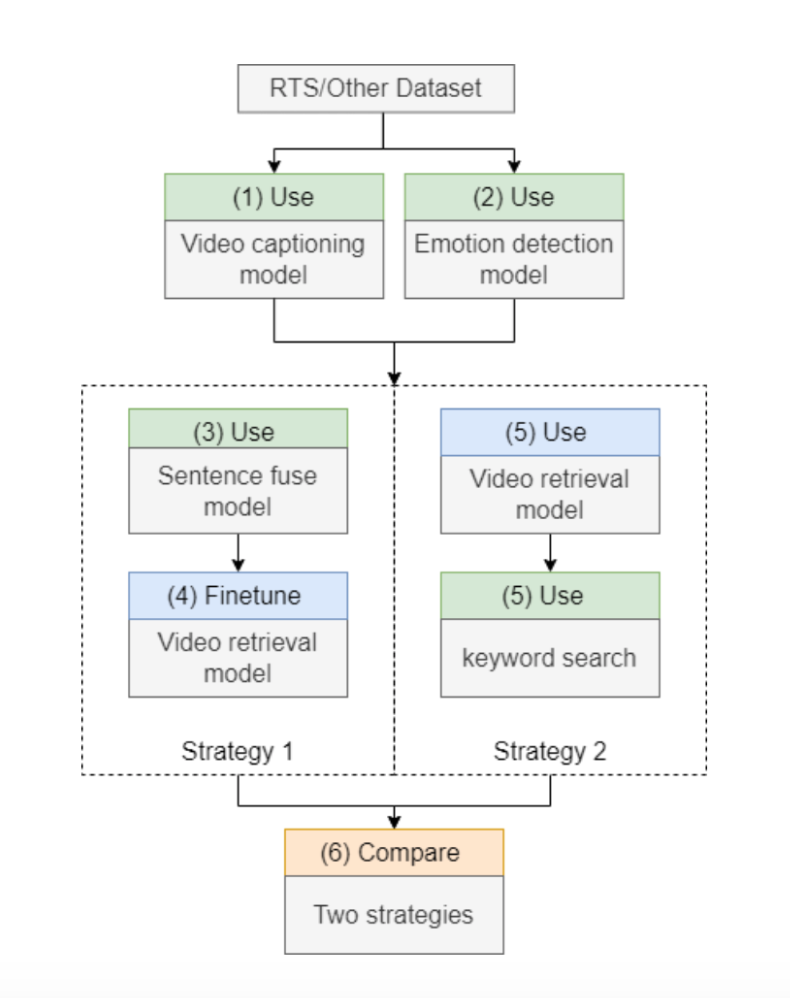} 
    \caption{The overall proposed pipeline structure}
\label{3}
\end{figure}

\begin{figure}
    \centering 
    \includegraphics[width=.8\columnwidth]{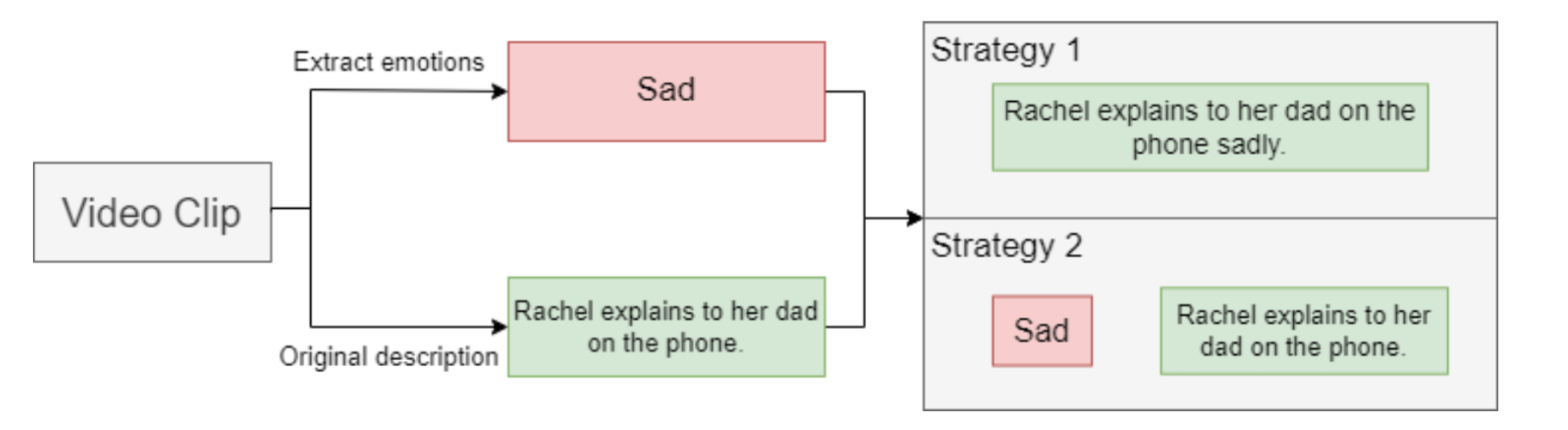} 
    \caption{An overview of two strategies to deal with queries with emotions.}
\label{4}
\end{figure}

\subsection{Video Segmentation}
We choose shot-based segmentations over speech-guided segmentation since the videos are all in Swiss French and are hard for transcription services to pick up correctly. We test three different shot-based segmentation methods: TransNet V2 \cite{souvcek2020transnet}, a content-aware detection that works by locating jump cuts, and adaptive content detection. The last two methods are wrapped by the PySceneDetect library, which is based on OpenCV. We resort to adaptive content detection since it performs the best based on our random sampling and qualitative evaluation. 

\subsection{Video Captioning}
We tried the newly available GPT-powered captioning models \cite{tewel2022zero}. Although gaining the most attention recently, model hallucinations of the GPT-powered ones introduce factual elements to the content - after processing a video of a math class, the model placed too much emphasis on the teacher and generated discussion about his religion and race. Due to this issue, we have not implemented such models.

Dense video captioning models give more fine-grained descriptions for each part of the video. To be more specific, compared to naïve models, they can generate multiple captions for a video. For example, a dense video captioning model can break a ten-second video into some events and describe them as “a man is looking at the mirror”, “a man is brushing his teeth”, and “a man in a white shirt is shaving” separately. 

We used the dense captioning model PDVC \cite{wang2021end} after initial experiments. It uses an event counter on top of a transformer decoder to take the context of the events into account, generating coherent captions. From the perspective of content coverage, PDVC is more suitable than naïve captioning models because it captures more events. In practice, PDVC generates ten captions for each video clip with timestamps. 

\subsection{Emotion Analysis}
The purpose of including the emotional analysis is twofold: 1) to test and verify if adding the emotion factor in the text-to-video retrieval model's training text-video pairs would yield a better and more comprehensive encoding for handling more complex query; 2) to prepare the dataset with emotional tags as metadata to test if a two-stage retrieval is better at handling emotionally abundant queries.

One of the most used emotion categorizations is Ekman's six basic emotions \cite{ekman1992argument}, which include happiness (joy), sadness, anger, disgust, surprise, and fear. Ekman believes these six universal emotions are the basic elements for more complex emotions from biological and evolutional perspectives. Many other emotion categorizations derive more emotions based on Ekman’s model, such as \cite{plutchik2001nature} and \cite{athar2011fuzzy}, which blend low-level emotions to obtain more complex emotions. For this project's scope, we follow Ekman’s categorization and test different models that can distinguish more than six emotions.
One challenge we face is that most models focus on classifying models based on facial expressions or speeches. Many datasets like CMU-MOSEI \cite{zadeh2018multimodal} only record the speaker’s voice and facial expressions, which makes the trained model inaccurate in real-world scenarios, and the speaker’s voice is mainly English, while our RTS dataset is in French. 

To tackle this issue, we resorted to a textless vision-language Transformer TVLT \cite{tang2022tvlt} for this task, which does not use text-specific modules such as tokenization or automatic speech recognition (ASR). 

Another challenge is determining neutral emotions. In almost all standard datasets, and as a result, in the emotion detection models, happiness (or joy) is the most common emotion, as less intense emotions are annotated and hence further classified as happiness. To fix the heavily unbalanced emotion distribution, we resort to a rule-based system to re-balance the distribution:
(1) If the highest probability score is not “happiness”, we assign the highest probability emotion to the clip.
(2) If the highest probability score is  “happiness” and the second-highest score is  “sadness”, we assign “happiness” to the clip (As it is unlikely for models to mistake sadness for happiness).
(3) If the highest probability score is  “happiness”  but the second-highest score is not  “sadness”, we assign the second-highest emotion to the clip (as it is very likely the less obvious emotion is miscategorized as happiness).

\subsection{Paraphrasing}
We post-processed the PDVC generateed descriptions. We eliminated duplicates and failed video descriptions and used the Parrot model \cite{damodaran2021parrot} to create paraphrased descriptions. 

Working on top of \cite{casas2021emotional}, we used the identified 35 emotion words as the seed words. Then, we ask three annotators to rewrite example sentences using these seed words and their derivatives as soon as possible, limiting them to adding new words only at the end of the sentences to simplify pattern summarization. Table \ref{t2} below shows all types of suffix patterns.

\begin{table}
\begin{tabular}{cl}
\multicolumn{1}{l}{\textbf{Emotion}} &
  \textbf{Suffix} \\ \hline
anger &
  \begin{tabular}[c]{@{}l@{}}in anger, with anger, angrily, in annoyance, \\ with annoyance, with hate, in disapproval\end{tabular} \\
disgust &
  in disgust, with disgust, disgustedly \\
happiness &
  \begin{tabular}[c]{@{}l@{}}with joy, joyously, joyfully, in amusement, \\ with amusement, with excitement, in excitement, \\ excitedly, with relief, with happiness, happily, \\ with enthusiasm, enthusiastically\end{tabular} \\
sadness &
  \begin{tabular}[c]{@{}l@{}}with sadness, sadly, in disappointment, \\ disappointedly, with grief, in grief, pessimistically\end{tabular} \\
fear &
  \begin{tabular}[c]{@{}l@{}}in fear, with fear, out of fear, fearfully, \\ from nervousness, out of nervousness, nervously, \\ with worry, worriedly, confusedly\end{tabular} \\
surprise &
  \begin{tabular}[c]{@{}l@{}}in surprise, with surprise, surprisedly, \\ with curiosity, curiously\end{tabular}
\end{tabular}
\caption{All suffixes produced by annotators. These suffix patterns then apply to all video descriptions.}
\label{t2}
\end{table}

Using the summarized patterns of suffixes, we randomly add the corresponding suffix to each description based on the emotion classification of the video clip. Then, we apply the Parrot model again to generate more variations for the rewritten results with the same model parameters as the previous time. The sentences generated by the model have more diverse positions for emotional words and derive new emotional words such as “rage”. The sentences obtained through hard coding and the sentences generated by the Parrot model are added to the final dataset as emotional video captions. 

\subsection{Naïve and Emotional Dataset}
\begin{figure}
    \centering 
    \includegraphics[width=.8\columnwidth]{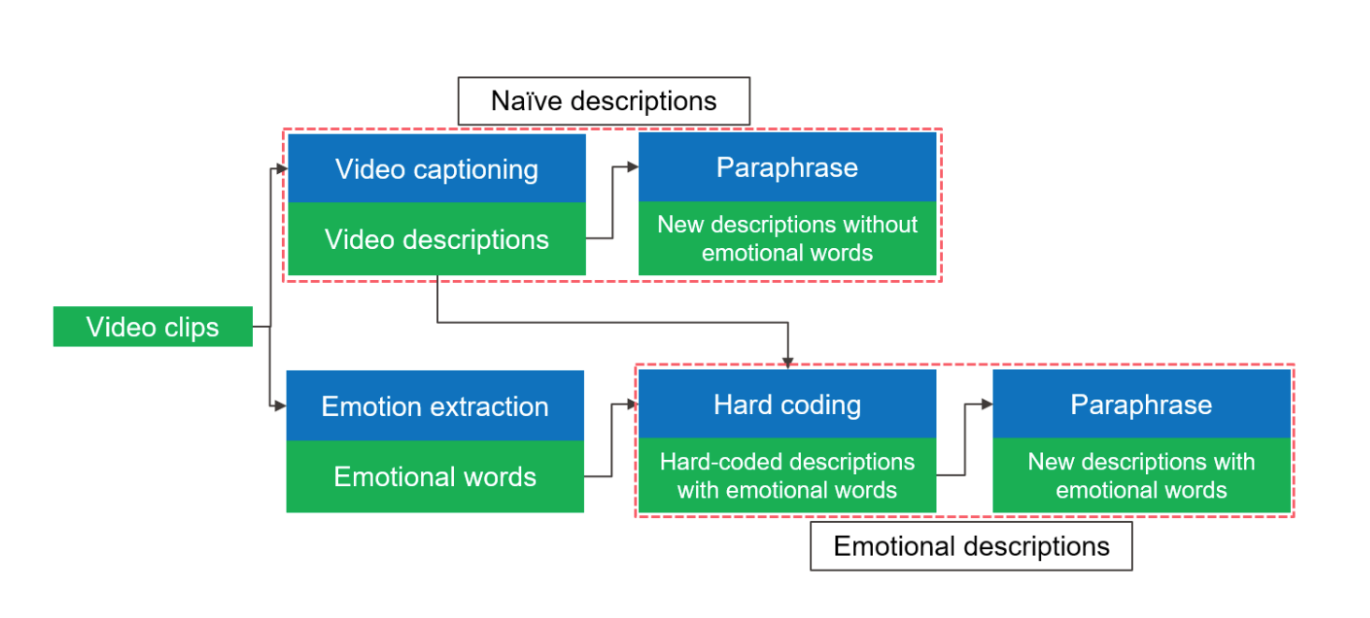} 
    \caption{Building process for naïve description set and emotional description set.}
\label{5}
\end{figure}

We obtained one dataset From the previous steps with naïve video descriptions and the other with emotional video descriptions. Both datasets share the same video clips. For simplicity, we will call them "naïve dataset" and "emotional dataset" separately. The Naïve dataset uses “naïve descriptions”, and the emotional dataset uses “emotional descriptions”. Figure \ref{5} concludes with details of how both description sets are built. 

\subsection{Text-to-video Retrieval}
Using these two datasets, we can finetune the baseline pre-trained text-to-video retrieval models verify if text-to-video retrieval models are suitable for our real-world dataset and conduct the proposed two strategies for handling more complex queries with emotion.
The number of naïve descriptions and emotional descriptions for the same video clips are usually not the same, which may affect fine-tuning results. Therefore, for each video clip, we randomly select the same number of descriptions from the larger description set according to that of the other set. Finally, the description set was processed into the same format as the standard dataset MSR-VTT.
The evaluation metric for text-to-video retrieval is "R@K", meaning the proportion of test queries that find the correct clips in the top K results. For example, R@10 = 0.4 means there are 40\% of test queries have the correct video clips in their top 10 retrieval results. 

\section{Results}
\subsection{Video Segmentation}
\begin{figure}
    \centering 
    \includegraphics[width=.8\columnwidth]{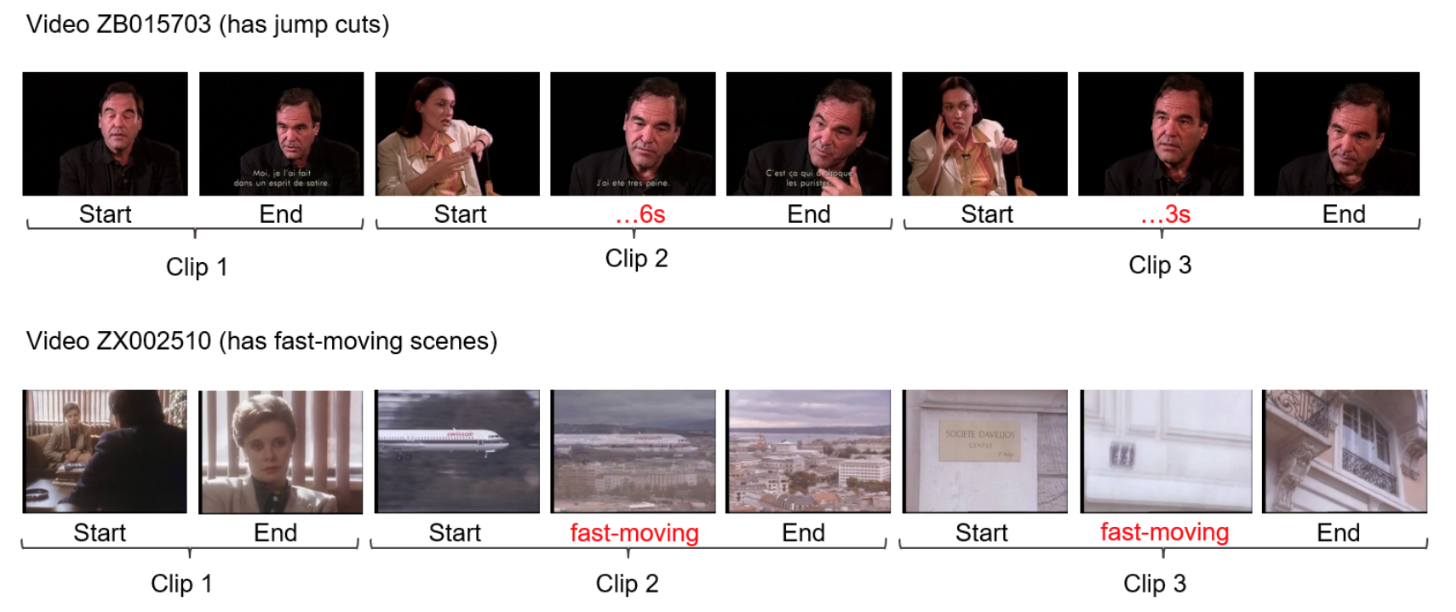} 
    \caption{Segmentation results by adaptive content detection.}
\label{6}
\end{figure}

As a result, we got 73,348 short clips from 816 videos. Of which 50\% are less than 18 seconds. Our proposed method also dealt with the two main issues regarding video segmentation. Figure \ref{6} shows two results of our video segmentation step. In the first case (Video ZB015703), clip \#2 starts with a talking woman. At 00:00:06, there is a “jump cut” where the camera switches to another person. The detection method regards this moment as the end of this clip, but as this clip does not reach 12 seconds for now, we merge it with the subsequent clips until the total duration exceeds 12 seconds. The same situation also happens in clip \#3.
The second case (Video ZX002510) has fast-moving scenes in some clips. The detection method is not sensitive to such scenes and does not cut them into shorter clips, preserving the completeness of the clips. 

\subsection{Video Captioning}
A random 1\% sample was collected and evaluated by a team of 3 annotators. Annotators were asked to reference the standard dataset samples from MSR-VTT and determine if the results from the captioning models were comparable or a failure. The annotator's evaluation yields an 83\% comparable rate, slightly lower than expected. However, considering many mistakes in the MSR-VTT dataset (such as grammar, spelling, and factual mistakes), we determined the quality of the generated captions usable.
\begin{figure}
    \centering 
    \includegraphics[width=.8\columnwidth]{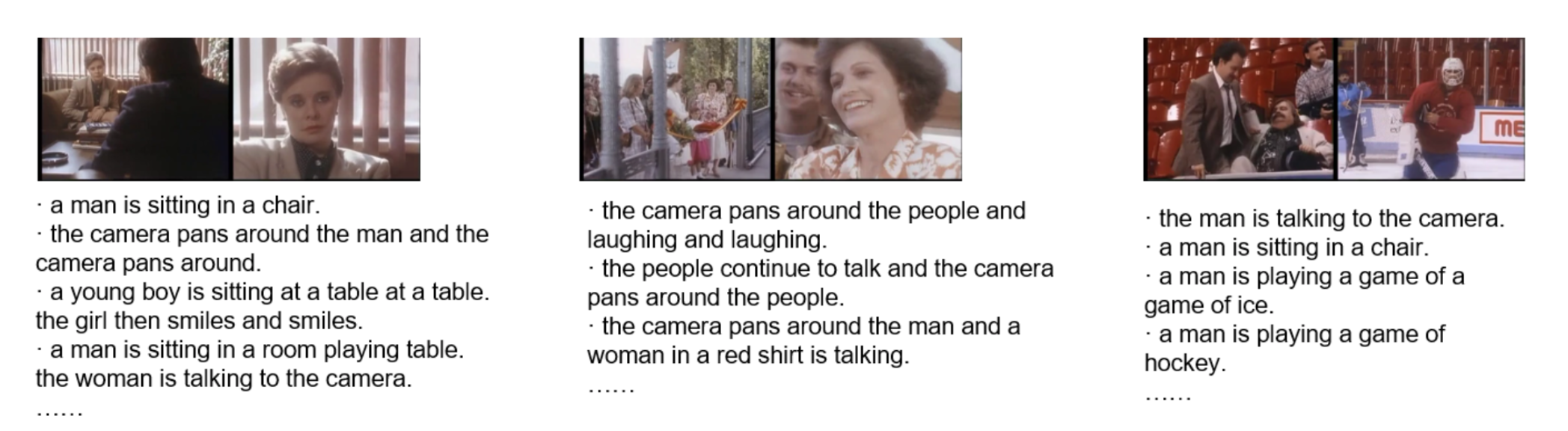} 
    \caption{Generated video description examples by PDVC model.}
\label{7}
\end{figure}

\subsection{Emotion Analysis}
As expected, the detected emotions are highly unbalanced, with more than 87.4\% recognised as happiness. After our rule-based reassign method, the distribution is much more balanced, with "happiness" dropping to 65.6\%. A similar evaluation is conducted for video captioning results using the balanced annotation. The annotator's evaluation yields a 76\% acceptance rate.

\subsection{Paraphrasing}
\begin{figure}
    \centering 
    \includegraphics[width=.8\columnwidth]{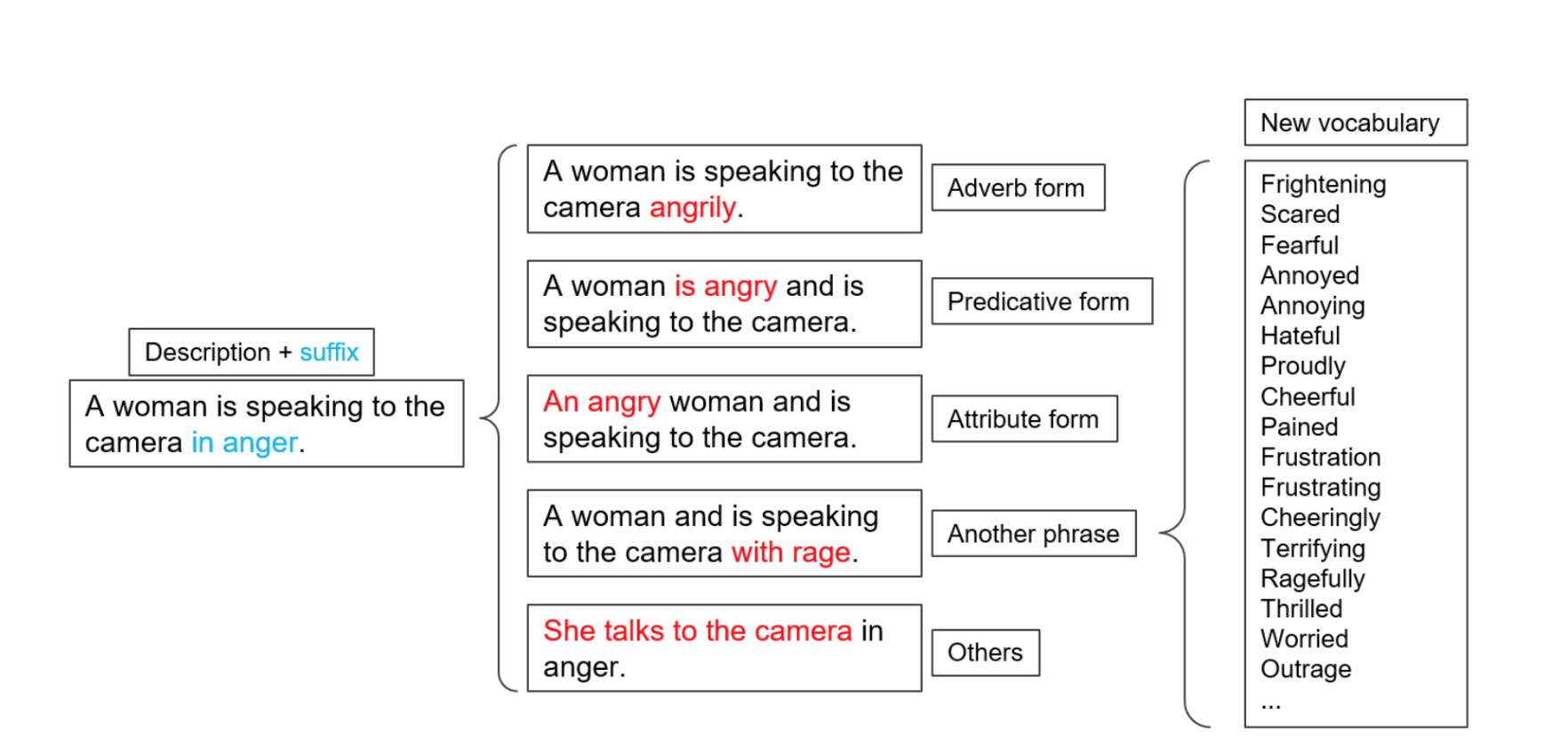} 
    \caption{Examples of paraphrased sentences generated by the Parrot model.}
\label{8}
\end{figure}

Figure \ref{8} shows a summary of an example of the paraphrasing and fusing of emotional words. One of the great things about using such a model to fuse the emotional words is that it automatically changes the suffix into other variants, adjusts the sentence structure, and introduces new emotion-related words that do not appear in our predefined 35 seed words. Using this, we obtained a new emotional dataset matching video clips with emotional descriptions. 

\subsection{Text-to-video Retrieval}
After post-processing, the two datasets we used are of the same properties as the standard dataset used for such tasks. Table \ref{t4} compares the naïve, emotional, and MSR-VTT datasets.

\begin{table}
\begin{tabular}{lccc}
 & \multicolumn{1}{l}{\textbf{Naïve}} & \multicolumn{1}{l}{\textbf{Emotional}} & \multicolumn{1}{l}{\textbf{MSR-VTT}} \\
\#captions / video & 13.11       & 9.82        & 20    \\
\#words / caption  & 9.60        & 11.72       & 9.28  \\
Avg. duration(s)   & \multicolumn{2}{c}{26.01} & 14.88
\end{tabular}
\caption{Comparison of the three datasets.}
\label{t4}
\end{table} 

For both datasets, we used 90\% of the clips as the training data and 10\% as the validation data. For the validation set, we only keep one description at random. We used a pre-trained baseline model CLIP4clip \cite{luo2022clip4clip} and a pre-trained improved model TS2-Net \cite{liu2022ts2} for evaluating the performance of text-to-video retrieval performances on the RTS dataset. Each of the pre-trained models is fine-tuned on our naïve and emotional datasets using the default training setting for both models, resulting in the naïve and emotional models separately. In the evaluation process, we evaluate each fine-tuned model on both types of descriptions using the two strategies proposed in Figure \ref{3}. The quantitative results are shown in Table \ref{t6}. 

\begin{table}
\begin{tabular}{llllll}
\multicolumn{1}{l}{\textbf{}} &
  \multicolumn{1}{l}{\textbf{Model}} &
  \textbf{Eval on} &
  \multicolumn{1}{l}{\textbf{R@1}} &
  \multicolumn{1}{l}{\textbf{R@5}} &
  \multicolumn{1}{l}{\textbf{R@10}} \\ \cline{2-6} 
  \textbf{CLIP4Clip} & Zero-shot  & Naïve     & 8.6          & 10.6           & 16.4\\
                     &            & Emotional & \textbf{9.8} & \textbf{11.1}  & 
 \textbf{17.7}  \\
                    & Naïve       & Naïve     & \textbf{23.0} & \textbf{33.1} & 
 \textbf{41.4} \\
                    &             & Emotional & -            & -             & - 
 \\
                    & Emotional    & Naïve     & -            & -             & - 
 \\
                    &             & Emotional & \textbf{24.5} & \textbf{36.3} & 
 \textbf{45.4} \\ 
 \cline{2-6}  
 \textbf{TS2-Net}   & Zero-shot   & Naïve     & 6.6          & 9.7           & 15.4\\
                    &             & Emotional & 8.9          & 10.0           & 
 16.8\\
                    & Naïve       & Naïve     & \textbf{26.6} & \textbf{37.9} & 
 \textbf{45.6} \\
                    &             & Emotional & 20.5          & 29.7          & 
 38.4\\
                    & Emotional   & Naïve     & 32.2          & 44.7          & 
 49.6\\
                    &             & Emotional & \textbf{32.4} & \textbf{46.1} & 
 \textbf{50.4}
\end{tabular}
\caption{Evaluation results of different models.}
\label{t6}
\end{table}

It is noted that these metrics are significantly lower than the average metrics of mainstream video retrieval models. We believe that there are several reasons for this: 

(1) Dataset. The video descriptions generated by PDVC models are repetitive, that is, many video clips may have a “talking to the camera” moment. However, in the model evaluation process, the videos are one-to-one and corresponded with the captions. It is quite common that although the model finds a video consistent with the caption, the match is still counted as a failure case because such a match is not recorded as the ground truth. Moreover, the inconsistency of video content may lead to the semantic inconsistency of video descriptions. A video clip can consist of multiple consecutive shots and may talk about more than one event, making PDVC describe multiple events for a single video clip. However, in some mainstream video retrieval datasets, the videos are usually carefully segmented so the content is more concentrated. Their manually written descriptions all describe the same thing with clear correspondence, making the one-on-one evaluation more probable.

(2) Training settings. Researchers often pre-train the models on larger text-to-video retrieval datasets such as WebVid2.5M, which can be expensive for our experiment environments. Both models perform much better than zero-shot methods. This means that fine-tuning the target datasets is necessary for text-to-video retrieval tasks. These comparisons also verify the quality of our data. Our generated descriptions, along with the paraphrased descriptions, do have some connections to the video contents. The models cannot archive such improvements if most descriptions are just random sentences.

\begin{figure}
    \centering 
    \includegraphics[width=.8\columnwidth]{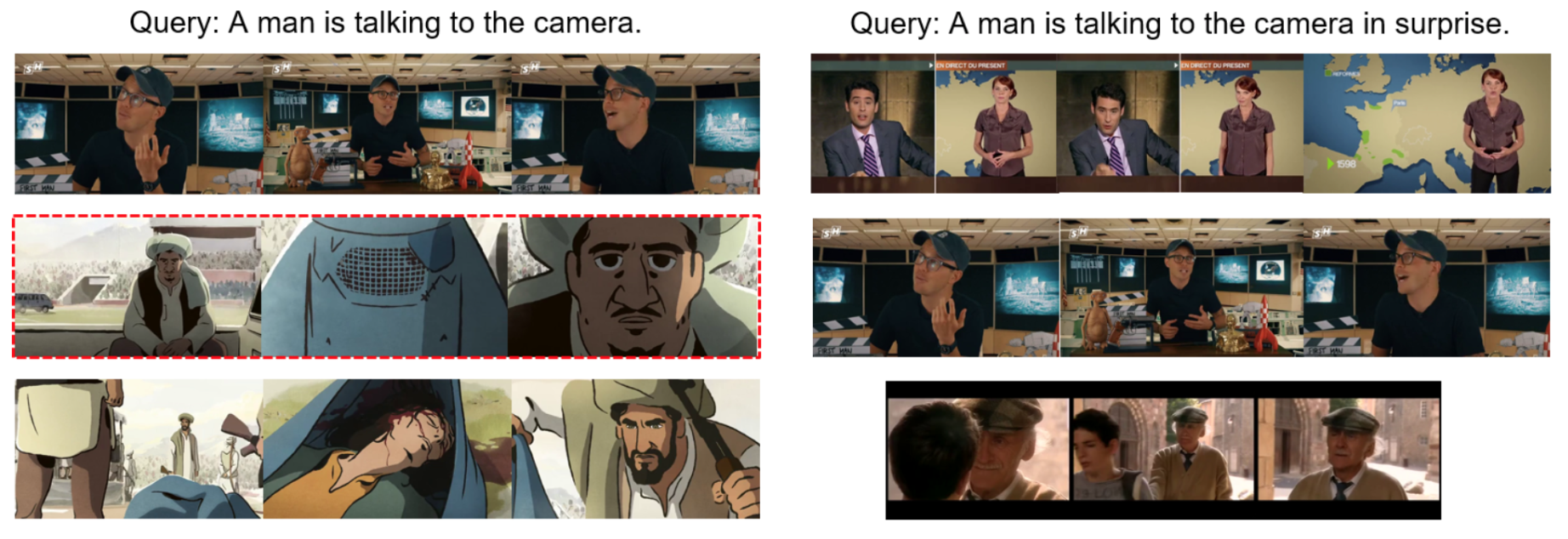} 
    \caption{Top 3 retrieved results for the first strategy (left) and second strategy (right).}
\label{9}
\end{figure}

Figure \ref{9} shows the retrieval results for two strategies. The original query for both strategies is “A man is talking to the camera in surprise”. 

Strategy 1 works by splitting the query into one emotional keyword binary search and one plain (without emotion) text-to-video retrieval task. 

Strategy 2 feeds the original query (“a man is talking to the camera in surprise”) to the emotional model.

Strategy 1 has less running time because the naïve model only runs on a specific emotion category. However, if the emotion detection is not accurate or even wrong for some video clips, strategy 1 suffers. In Figure 20, the second result of strategy 1 does not convey or contain the “surprise” element. Another disadvantage is that, although exacting emotion from queries is simple, simplifying emotional descriptions to naïve descriptions is tricky and difficult. It is overwhelming to summarize all the simplifying patterns, and no efficient tool can paraphrase sentences with certain orientations. 

Strategy 2 runs longer because it runs a full similarity search on all vectors with the query embedding rather than a subset of the videos. But it also means this method can expend or connect more serendipitously connected contents. And it is simpler when it comes to preparing the dataset.

Based on the experiments, we verified that our pipeline and strategies are effective in 1)pre-processing real-world archival content to adapt the text-to-video encoding. 2)Automatically generate and augment text descriptions of video clips that could be used to handle more complex natural language queries (such as fused with emotional words).

\section{Application the Latent Wanderer}
\subsection{Motivation}
Although the pipeline works for text-to-video retrieval tasks, the results are not accurate enough to serve real-life retrieval tasks if the purpose is to find a specific video. However, the pipeline is a great exploration method because it intuitively builds unexpected connections among archival content. Most interfaces and applications focusing on search cannot provide users with such an explorative experience. 

Based on these points, we propose and design an exploration interface prototype on the top of the text-to-video retrieval model, allowing users to browse videos better and understand the latent space of text-to-video retrieval models. The ideal setting for the interface would be a VR environment with a user localization function. 

\subsection{Design}
To make the interface easy to interact with, we combine the ideas of (1) dimension reduction techniques in the machine learning field and (2) procedural board map generation in video games.
We use UMAP, and its official \cite{mcinnes2018umap} library to process our data since it is more suitable for non-continuous data. To enhance the enjoyment of the exploration experience, we tile different square grids to produce a game-like map, appealing to users to step into each grid and discover what will happen. 

Assume we already have a text-to-video retrieval model and a bunch of processed RTS video clips, we prepare the video data and the game map through the following steps:

(1) Input the video clips to the retrieval model via video encoder and get a bunch of high-dimensional vectors.

(2) Apply the UMAP algorithm to the high-dimensional vectors and convert them to a bunch of 2D vectors. 

(3) Treat the 2D vectors as 2D points in the game map and do subsampling, as shown in Figure 22. A negative grid (blue) will become positive (orange) if any 2D points are projected into its region.

\begin{figure}
    \centering 
    \includegraphics[width=.8\columnwidth]{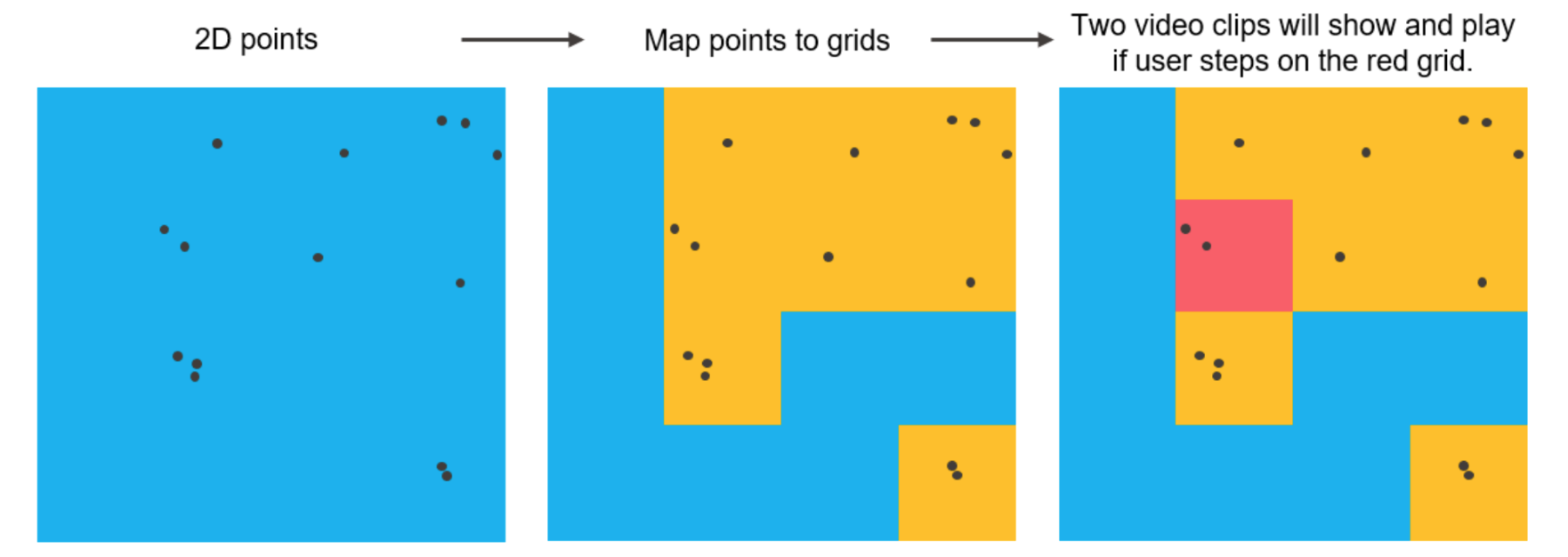} 
    \caption{Top view of the subsampling process, which projects and subsamples the 2D vectors into grids.}
\label{10}
\end{figure}

(4) If a user steps into a positive grid (for example, the red grid, which contains two points in Figure \ref{10}), the video clips corresponding to the two 2D points will show around the user and start to play.  

\begin{figure}
    \centering 
    \includegraphics[width=.8\columnwidth]{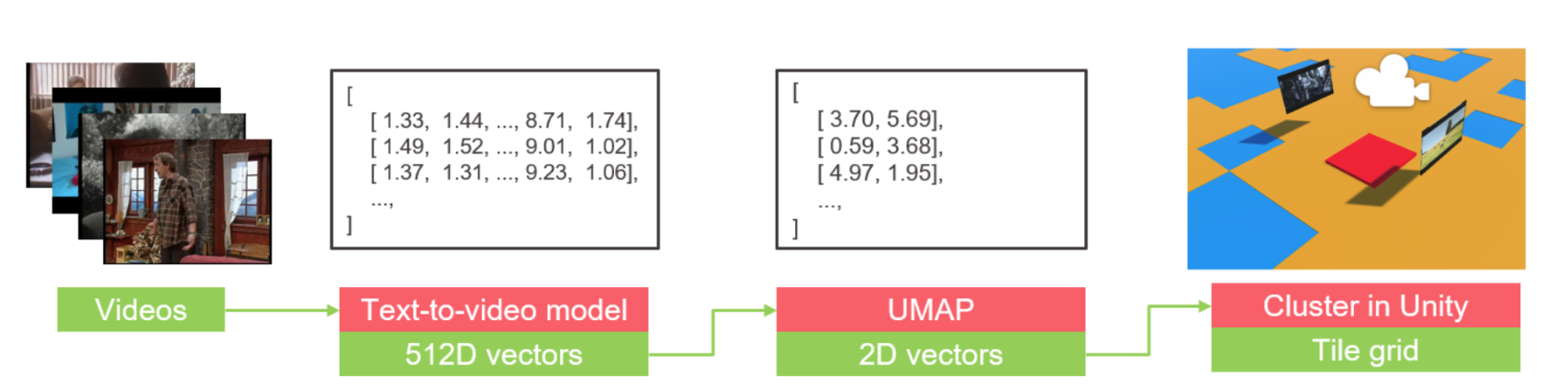} 
    \caption{Whole process of data preparation.}
\label{11}
\end{figure}

Figure \ref{11} shows the whole process of data preparation. In the last step, the red grid highlights where the user stands. It is not necessary to be another colour. Users can walk to neighbour grids to explore similar videos. 

\begin{figure}
    \centering 
    \includegraphics[width=.8\columnwidth]{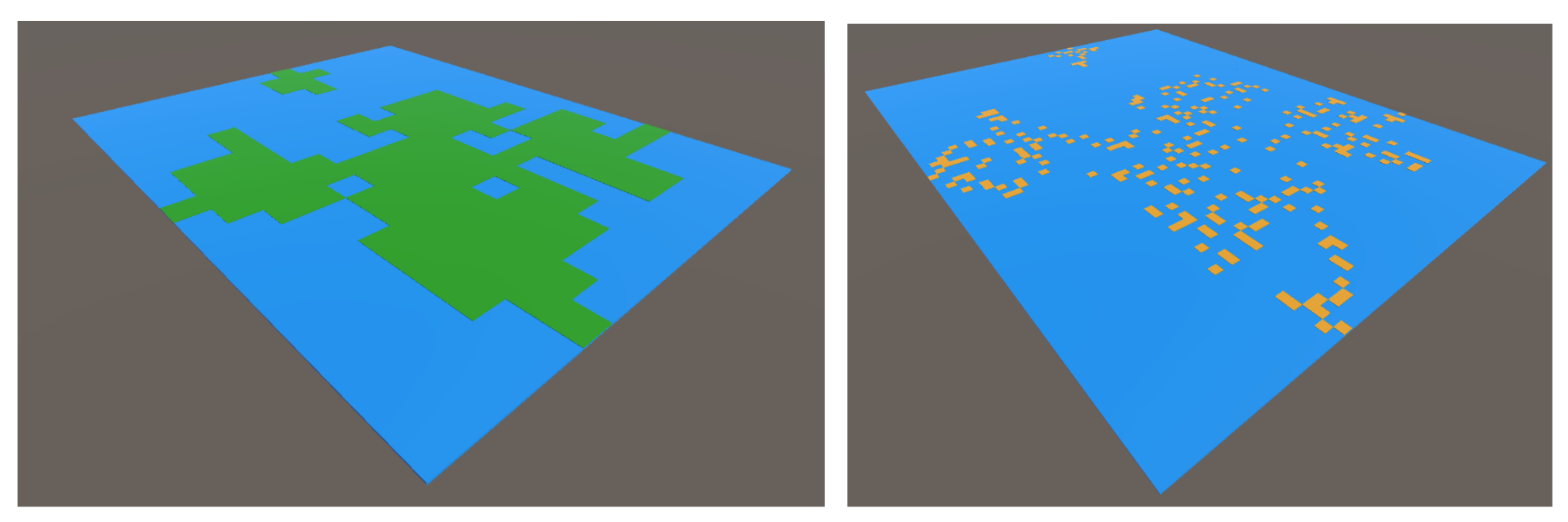} 
    \caption{An overview of two example maps.}
\label{12}
\end{figure}

The configuration of the map generation is also highly customized. Figure \ref{12} shows two example maps. Developers can change the colours and textures of both types of grids to archive cartoon or realistic styles, adjust the sizes of grids to generate a more fine-grained map and place any art assets to make the map more vivid. 

We propose that the latent wanderer illustrate the latent space with the holistic text and video joint embedding. Unlike previous latent space, which mostly focuses on one aspect such as colour, movement, or object detection, this latent combines many modalities and natural language descriptions. It served as an intuitive visualization of what machine learning does and allowed users to explore the entirety of the archive through a “natural language” perspective. Each point on the map is translatable to either corresponding sentences or the nearest videos. By travelling through islands or randomly walking, the user travels in a holistic semantic space. By comparing and connecting different places on the map, the user can explore and make sense of the archival content on their terms.

\section{Limitations and Future work}
Although we provide a complete pipeline from RTS raw data to a usable video retrieval framework, the work is based on a small-scale subset of the entire archive. The pre-training and fine-tuning are also at a miniature scale. It would be ideal to test the whole thing at a larger scale.

The pipeline can be functionally divided into independent modules. Each module's effect depends on the output quality of the previous modules. It would be ideal to have mode-tailored training sets for each step. 

The one-on-one evaluation for the text-to-video retrieval model cannot reflect real-world situations. It mostly relies on subjective observations to see if a retrieved item makes sense.

For each step in the pipeline, we present our reflections as follows:
The currently selected shot-based video segmentation method is relatively simple. It only considers visual elements, while the speech-based segmentation method ignores them, so the two methods can be combined. And the RTS data is very diverse. Ideally, tailored segmentation methods should be selected according to the type of program. 

When generating video descriptions, we can try to replace the CLIP modules used by PDVC with more advanced ones for better performance. The current model does not consider audio, so it will be promising to train a captioning model that considers that.

We only assign one emotion to each video, but a video may have complex emotions, and different characters may also have different emotions. A more fine-grained and complex emotion classifier is very much needed.

\section{Conclusion}
Advancing the trend of building explorative and interactive interfaces for cultural and heritage archives and public institutions to engage and realise the potential of content fully, this project achieves the following goals:

1) Design and verify a working pipeline to pre-process real-world archival videos to adapt the most recently developed text-to-video encoding. 

2) Verify that using the text-to-video retrieval modes can go beyond visually focused queries and simple and bland descriptions and handle more complex natural language queries. By analysing the performance of the two strategies, we can provide an improved strategy with minimum change to the current pipeline.

3) As an exemplary application for utilising such a higher-level encoding of videos, this work proposed a unity-based interactive latent map called "latent wanderer". This serves as an exploration interface prototype on stop of the text-to-video retrieval model, allowing users to browse videos better and understand the latent space of text-to-video retrieval models.



\newpage
\bibliographystyle{ACM-Reference-Format}
\balance
\bibliography{sample-base}

\appendix

\end{document}